\documentclass[prl,twocolumn,showpacs]{revtex4}

\usepackage{amsmath}
\usepackage{amssymb}
\usepackage{epsfig}

\begin{document}
\title{Memory erasure in small systems}
\author{Raoul Dillenschneider and Eric Lutz}
\affiliation{Department of Physics, University of Augsburg, D-86135 Augsburg, Germany}

\begin{abstract}
We consider an overdamped nanoparticle in a driven double--well potential as a generic model of an erasable one--bit memory. We study in detail  the statistics of the heat dissipated during an erasure process and show that full erasure may be achieved by dissipating less heat than the Landauer bound. We quantify the occurrence of such events and propose a single--particle experiment to verify our predictions. Our results show that Landauer's principle has to be generalized at the nanoscale to accommodate heat fluctuations.
\end{abstract}


\maketitle

Maxwell's demon is a salient figure of thermodynamics \cite{lef03}. Introduced in 1867 to illustrate the statistical nature of the second law, the demon is an intelligent microscopic  being able to monitor individual molecules contained in two neighboring chambers initially at the same temperature. By opening and closing a small hole in the separating wall, the demon collects the faster ('hot') molecules in one of the chambers and the slower ('cold') ones in the other, thus creating a temperature difference. The demon is therefore able to decrease the entropy of the system without performing any work, in apparent violation of the second law of thermodynamics. The paradox was eventually resolved by Bennett \cite{ben82}, who noted that during a full thermodynamic cycle, the memory of the demon, used   to record the coordinates of each molecule, has to be reset to its initial state. According to Landauer's principle \cite{Landauer}, memory erasure necessarily requires dissipation of entropy: the cost of erasing one bit of information is  at least $\Delta S_{Landauer} = k \ln 2$, where $k$ is the Boltzmann constant \cite{Landauer,Shizume,Piechocinka,kaw07}. The  entropy cost to discard the  information obtained about  each gas molecule appears to always exceed the entropy reduction achieved by the demon. Maxwell's demon   is hence exorcised by Landauer's erasure principle.    

The second law   stipulates that  irreversible entropy production is  positive in macroscopic systems. Thermal fluctuations are usually exceedingly small at these large scales and are therefore discarded. By contrast,  fluctuations become   predominant in microscopic systems and it has lately been recognized that the second law has  to be generalized to properly take positive as well as negative entropy fluctuations into account \cite{bus05}. This generalization  takes the form of a fluctuation theorem, $P(-\Sigma) = P(\Sigma) \exp(-\Sigma)$, for the probability distribution of the entropy production $P(\Sigma)$ \cite{Evans,gal05}. Processes with negative entropy production can  hence occur in small systems with small $\Sigma$, while they are exponentially suppressed in large systems with large $\Sigma$. The fluctuation theorem has recently been confirmed experimentally using a colloidal particle in a modulated optical trap \cite{car04} and a driven torsion pendulum \cite{dou06}.

Motivated by the recent development of  nanotechnological memory devices \cite{bad04,via06,jan08}, we investigate  
  the impact of fluctuations on memory erasure in small systems. By considering a generic  system, we show  that full  erasure may be achieved in nanosystems with an  entropy dissipation less than the Landauer limit. We  express the Landauer bound in terms of the free energy difference of the erasure process and quantify the probability of having a dissipated entropy below that value. We moreover discuss an experimental set--up where our findings
 can be  tested using currently available technology. To our knowledge, Landauer's principle has never been  verified experimentally, despite its fundamental importance as a bridge between information theory and physics. One of the main difficulty is that in order to access the dissipated entropy,  one needs to be able, like Maxwell's demon, to follow individual particles. Thanks to  recent progress of single molecule experiments, this  is now possible \cite{coh05,ACohen}.

\emph{The model.}
Following the original work of  Landauer, we consider an overdamped Brownian particle in a one--dimensional double--well potential as a generic model of a one--bit memory \cite{Landauer}.  The state of the memory is assigned the value {\it zero} if the particle is in the left ($x < 0$) well and {\it one} in the right ($x >0$) well. The memory is said to be erased when its state is reset to {\it one} (or alternatively {\it  zero}) irrespective of its initial state. The potential barrier is  assumed to be much larger than the thermal energy so that the memory can be considered stable   in the absence of any perturbation. We describe the dynamics of the Brownian  particle using the Langevin equation,
\begin{equation}
\gamma \frac{d x}{dt} = -\frac{\partial V}{\partial x} +  A f(t) 
+ \xi(t) \ , 
\label{Eq1}
\end{equation}
\noindent
where $\gamma$ is the friction coefficient and $\xi(t)$  a delta correlated Gaussian noise force with    
$\langle \xi(t)\rangle = 0$ and 
$\langle \xi(t) \xi(t^{'})\rangle = 2 D \,\delta (t-t^{'})$. The  diffusion coefficient is given by $D=\gamma k T$.
The double--well potential $V(x)$  is of the standard form, $V(x)=- a g(t)\, x^2/2 +
b \,x^4/4$, with a barrier height that can be  controlled by the dimensionless function $g(t)$.  The Brownian particle is additionally subjected to a driving force  $ A f(t)$ with  a driving amplitude $A$. 
In order to simplify the analysis of Eq.~\eqref{Eq1}, it is convenient to introduce dimensionless variables defined as $\bar{x} = x/x_m$ and $\bar{t} = a t /\gamma$ where $x_m  = \sqrt{a/b}$ is the position of the (positive) minimum of the potential
\cite{Hanggi}. The dimensionless potential is then $\bar{V}(\bar x) = - g(\bar t) \bar{x}^2 /2 + \bar{x}^4 /4$ and the rescaled driving amplitude and diffusion coefficient are respectively given by $\bar{A} = A/ a x_m$ and $\bar{D} = D/ \gamma a x_m^2$.
In the following, we will drop  the bar signs and consider  dimensionless quantities.

\begin{figure}
\center
\epsfig{file=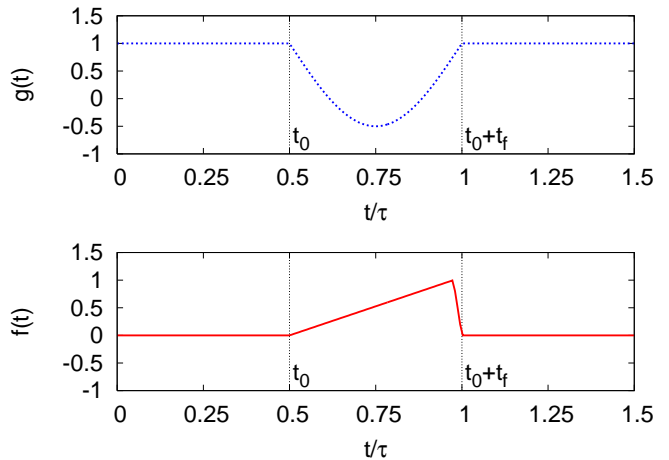,width=9cm}
\caption{(Color online) Time dependence of the two dimensionless functions  $g(t)$ and $f(t)$, Eqs.~\eqref{eq2} and \eqref{eq3}, that govern the erasure protocol. The function $g(t)$ lowers the potential barrier, whereas $f(t)$ induces a tilt that brings the Brownian particle into the right well (state {\it one} of the memory).}
\label{fig1}
\end{figure}

 As discussed in detail by Bennett in Ref.~\cite{ben82}, the memory can  be reset to {\it one} by {\it i}) lowering the barrier height and {\it ii}) applying an external tilt that brings the nanoparticle into the right well. We here characterize the erasure protocol with the help of the two functions $g(t)$ and $f(t)$ shown in Fig.~\ref{fig1}. The function $g(t)$ defined as,   
\begin{eqnarray}
\label{eq2}
g(t) =
\begin{cases}
& 1-C \sin( \omega (t-t_0)) \text{ if } t \in \left[t_0,t_0+t_f \right], \\
& 1 \text{ otherwise}.
\end{cases}
\end{eqnarray}
\noindent
lowers and raises  the potential barrier in a time $t_f = \tau/2$ specified by the period, $\tau = 2\pi /\omega$, of the sine function. The parameter $C$ controls the 
amplitude of the barrier lowering. On the other hand,
the tilting function $f(t)$  has   a sawtooth shape parameterized as,
\begin{eqnarray}
\label{eq3}
f(t) &=&
\begin{cases}
& (t-t_0)/\tau_1 \text{ if }
 t \in \left[ t_0 , t_0 + \tau_1 \right],  \\
& 1-(t-t_0-\tau_1)/\tau_2 \\
& \hspace{2cm} \text{ if } t \in \left[ t_0+\tau_1, t_0 + t_f \right], \\
& 0 \text{ otherwise} \ .
\end{cases}
\end{eqnarray}
\noindent
The two time constants $\tau_1$ and $\tau_2$ verify $t_f = \tau_1 + \tau_2$
and are respectively the times during which the driving  force $A f$ is ramped up 
to its maximal value $A$ and then down again to zero. The values of the parameters in Eqs.~\eqref{eq2} and \eqref{eq3} are chosen in order to minimize the dissipated entropy. After a full erasure cycle, both $g(t)$ and $f(t)$ take back  their original values and the double--well potential $V(x)$ is restored to its initial shape. 

The energetics of the Brownian particle can be  introduced by following the prescription of Ref.~\cite{Sekimoto}. The variation of the total potential energy  is  defined as $\Delta U = U(x(t_0+t_f),t_0+t_f) - U(x(t_0),t_0)$, where $U(x,t) = V(x,t) - x A f(t)$ is the sum of the double--well potential $V(x,t)$ and the driving potential $-x A f(t)$. The  work performed on the particle is in turn given by,  
\begin{equation}
\label{eq4}
W = \int_{t_0}^{t_0+t_f} dt \,\frac{\partial U(x,t)}{\partial t} \ .
\end{equation}
Since the driving force satisfies $f(t_0+t_f)=f(t_0)=0$, the work can be  rewritten in terms of the functions $g(t)$ and $f(t)$ as $W = \int_{t_0}^{t_0+t_f} dt\,
\dot x [A f(t) + (g(t)-1) x ]$. This expression indicates that particles  moving against the total force,
$F(x,t) = A f(t) + (g(t)-1) x $, will generate work, while particles moving  in the same direction  will absorb work. This is the physical origin of work fluctuations in the present system.
According to the first law, the heat dissipated into the bath is the difference between  work and  total energy change, $Q= W-\Delta U$. It is important to note that  the energy change during an erasure cycle vanishes on average, $\langle \Delta U\rangle = 0$. As a consequence, the mean work done on the system and the mean dissipated heat are equal, $\langle W \rangle = \langle Q \rangle$. Moreover,  the dissipated heat is related to the dissipated entropy via $\langle Q \rangle = T \Delta S$ for a quasistatic transformation. The Landauer bound for the dissipated heat then follows as $\langle Q \rangle_{Landauer} = kT \ln 2$, or equivalently, $\langle \bar Q \rangle_{Landauer} = \bar D \ln 2$ in dimensionless units.

\begin{figure}
\center
\epsfig{file=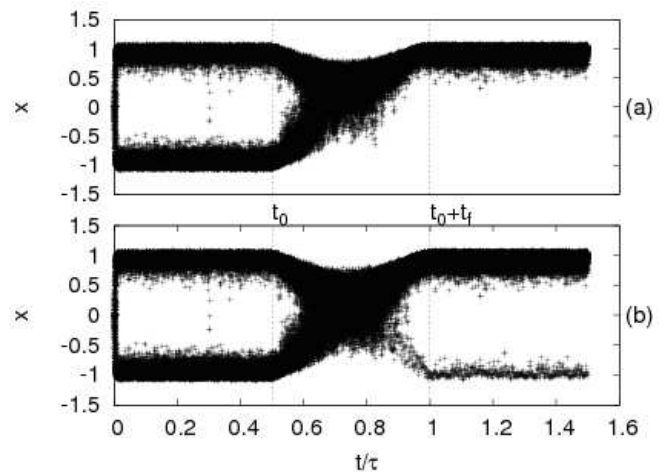,width=8.8cm}
\caption{Stochastic evolution of an ensemble of 100 trajectories during the erasure process. The state of the memory is {\it zero } or {\it one} with probability one half at initial time $t_0$. Erasure of the memory to state  {\it one} is complete in case (a), while only partial in case (b). We have here used  $\omega = 0.01$ and $D =0.02$.}
\label{fig2}
\end{figure}

\emph{Numerical results.}
We have numerically integrated the overdamped Langevin equation \eqref{Eq1}  using the Heun method 
(Runge-Kutta $n=2$) \cite{gre88} with a Gaussian noise force generated by means of the 
Box-Muller method \cite{NumericalRecipies}. In our  simulations, the particle is initially placed 
at position $x=0$ and is  then left to equilibrate with the heat bath at 
temperature $T$. After a thermalizing time $t_0$,  the memory is therefore either in state {\it zero} or {\it one} with probability one half. Two examples of an erasing sequence for different driving amplitudes (see below) are shown in Fig.~\ref{fig2} for  an ensemble of  $10^2$ particles. 
The erasure protocol is applied at $t=t_0$  for a duration of 
$t_f$. We observe that in the first case, all particles end up in the positive potential well. Full reset of the memory to state {\it one} is therefore achieved. On the contrary, only partial erasure is realized in the second case, as some Brownian particles remain in the negative potential well after the end of the erasure process.

\begin{figure}
\center
\epsfig{file=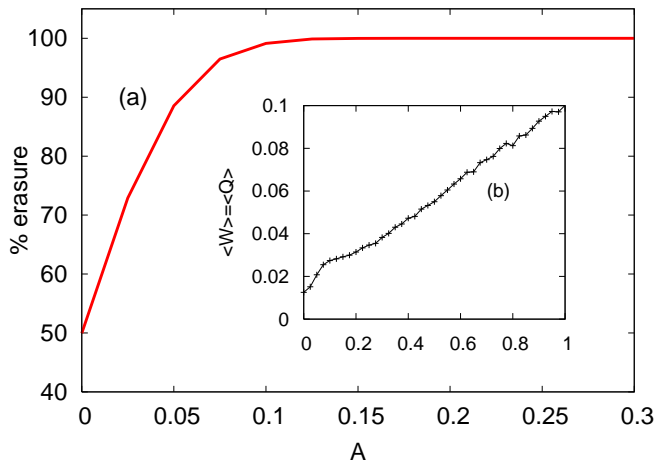,width=9.2cm}
\caption{(Color online) (a) Memory erasure rate, defined as the relative number of particles ending in the right potential well, as a function of the amplitude of the tilting force $A f(t)$; full memory erasure is attained for $A>A_0 \simeq 0.15$. (b) Mean work $\langle W\rangle $ and mean dissipated heat $\langle Q \rangle$ versus the tilting amplitude $A$ for parameters $\omega = 0.01$, $C=1.2$, 
$D=0.02$,  $\tau_0=0.05 t_f$ and
$\tau_1 = 0.45 t_f$.}
\label{fig3}
\end{figure}

Figure \ref{fig3} shows the erasure rate, defined as the relative number of particles being in the right potential well at the end of the erasure protocol, as a function of the driving amplitude $A$. The simulations are performed with $10^5$ trajectories and a time slicing $\Delta t = \tau*5.10^{-5}$. We note that full erasure is obtained for  driving amplitudes larger than a threshold value of $A_0\simeq 0.15$. The mean work $\langle W \rangle$ and the mean dissipated heat $\langle Q \rangle$ during an erasure cycle are plotted in the inset: both increase monotonically with increasing driving amplitude. For the chosen parameters, the dimensionless diffusion coefficient is $D= 0.02$, corresponding to a Landauer limit of $ \langle Q\rangle_{Landauer} = D \ln 2 = 0.014$. We stress  that  the mean dissipated heat always exceeds the Landauer bound, $\langle Q\rangle >\langle Q\rangle_{Landauer}$,  in accordance with   Landauer's principle. We have similarly studied the dependence of the erasure rate on the parameter $C$ (not shown).
 
\begin{figure}
\center
\epsfig{file=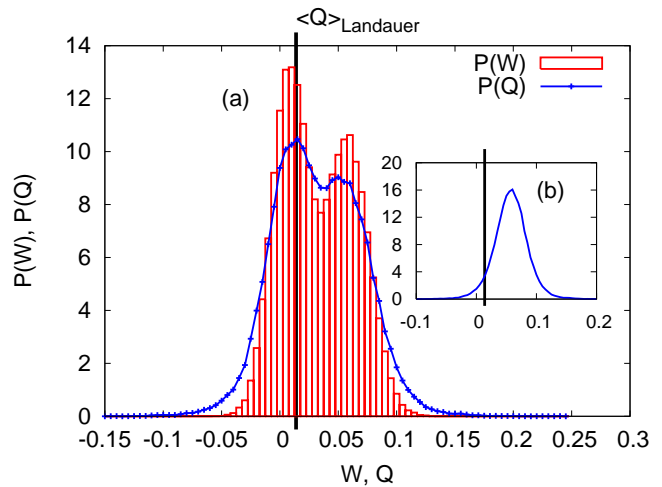,width=8.9cm}
\caption{(Color online) (a) Distribution of work and heat for a full erasure process. (b) Conditional heat distribution for particles initially in the left well. The existence of fluctuations below the Landauer limit is here clearly visible.
Parameters are  $\omega =0.01$, $A=0.2$, $C=1.0(a), 1.2(b)$, and $D=0.02$.  The average work and heat is  given by $\langle W \rangle = \langle Q \rangle =0.036$ and the Landauer bound is $ \langle Q\rangle_{Landauer} = D \ln 2 = 0.014$.}
\label{fig4}
\end{figure}

The full probability distributions  of work  and  heat  for a given 
protocol resulting in complete memory erasure are depicted in Fig.~\ref{fig4}.  Three important  points are worth discussing. First, $P(W)$ and $P(Q)$ are broad distributions with positive and negative fluctuations induced by  the thermal bath. We note furthermore that the heat distribution is   markedly broader than the work distribution. The difference in width of the two distributions  can be understood by noticing that, following the definition of $Q$, its variance is larger than the sum of the variances of work  and energy, $\sigma^2_Q\geq \sigma^2_W + \sigma^2_{\Delta U} > \sigma^2_W $. The second observation is that both distributions have  a bimodal structure. The left peak  arises from half the particles being initially located in the right potential well, 
whereas the right peak stems from  particles being moved from the left to the right potential well during the erasure process. The inset shows the conditional heat distribution pertaining to these particles.
Third, and most importantly, we note that a significant fraction of  trajectories lead to full memory erasure while dissipating less heat than the Landauer limit, $Q <\langle Q\rangle_{Landauer}$. From the conditional heat distribution shown in the inset,   we find that  $7.4 \%$ of all trajectories  starting in the left potential well at time $t_0$ yield a dissipated heat below $\langle Q\rangle_{Landauer}$. Full memory erasure in small systems can hence be obtained  {\it below} the  Landauer limit due to thermal fluctuations.

\emph{Analytical considerations.}
We  next  derive  the Landauer bound  for a nanoparticle in the double--well potential. Our starting point is the Jarzynski equality \cite{Jarzynski},
\begin{equation}
 \langle e^{-\beta W}\rangle = e^{-\beta \Delta F} = \frac{Z_1}{Z_0} \ ,
\end{equation}
which relates the average exponentiated work done on the Brownian particle to the free energy difference $\Delta F = -kT \ln Z_1/Z_0 $. Here $Z_0$ and $Z_1$ are the initial and final partition functions of the system. The Jarzynski equality is valid for arbitrary changes of the potential $U(x,t)$ and has been successfully verified in biomolecule pulling experiments \cite{lip05}. Before erasure,   particles  are thermally distributed over the entire double--well potential and the initial partition function is thus $Z_0 = \int_{-\infty}^\infty dx \exp(-\beta V(x))$. After successfull erasure, all particles are thermalized  in  the right potential well and $Z_1 = \int_0^\infty dx \exp(-\beta V(x))= Z_0/2$. As a result, the ratio of the partition functions is simply $Z_1/Z_0=1/2$ and $\Delta F = kT \ln 2$. 
By  using Jensen's inequality, 
$\langle \exp(-\beta W) \rangle \ge \exp(-\beta \langle W\rangle)$, we immediately obtain, $\langle W \rangle = \langle Q \rangle \ge k T \ln 2$. The Landauer bound  is thus given by   the free energy difference of the erasure process and is  only attained in the quasistatic limit when $\langle W \rangle = \langle Q \rangle =\Delta F$.  By further generalizing an argument presented in \cite{jar99}, the probability to observe a value of dissipated heat below the Landauer limit, $\langle Q \rangle_{Landauer} -q$, with $ q >0$ can be estimated to be,
\begin{equation}
 \mbox{Prob}[  Q <\langle Q \rangle_{Landauer} -q] \leq \exp(-\beta q) \ .
\end{equation}
Small fluctuations below the Landauer limit are hence possible. However, large fluctuations, $q\gg kT$, are exponentially suppressed, in agreement with the  macroscopic formulation of Landauer's principle. We can therefore conclude that at the nanoscale, where fluctuations cannot simply be neglected, Landauer's principle has to be generalized in a way similar to the second law.

\emph{Experimental implementation.} The control of single nanoparticles in arbitrary 
two--dimensional force fields has been demonstrated in Refs.~\cite{coh05, ACohen}. In these 
experiments,  fluorescence microscopy is combined with real--time feedback techniques to 
manipulate  nanoscale objects (from 50 to 200 nm) in solution via a position--dependent 
electrophoretic force. The force may be varied with high precision over nanometer distances 
and millisecond times. The trajectory of the particle is monitored with a high--sensitivity 
CCD camera. The Brownian motion of a 200 nm fluorescent polystyrene bead in a {\it static} 
double--well potential has been investigated in Ref.~\cite{ACohen} and the measured hoping 
rates between the two wells have been successfully compared with Kramers theory.
We propose to study memory erasure in nanosystems by extending the former experiment to a 
{\it driven} double--well potential according the erasure protocol discussed above. The work 
$W$ done on the particle as well as the dissipated heat $Q$ can be determined directly from 
the measured particle trajectory via Eq.~\eqref{eq4}.  By assuming a spacial diffusion coefficient 
$D_x = 10 \mu \mbox{m}^2 \,s^{-1}$, a measuring time  $\tau = 60 s$ and a temperature  $T=300 K$, 
we obtain the following realistic values for the parameters of the Langevin equation 
\eqref{Eq1}:  
$\gamma = k T/D_x \simeq 4.10^{-10} \, kg\,s^{-1}$,
 $a = \gamma \bar{\tau}/\tau 
\simeq 4.10^{-9} \, kg \,s^{-2}$ and 
$b = \bar{D} a^2/k T \simeq 40 \, kg\,m^{-2}\,s^{-2}$ with   a dimensionless execution time 
$\bar{\tau} = 600$ and diffusion coefficient $\bar{D} = 10^{-2}$, as in our numerical simulations.
The minima of the potential  are then separated by $\Delta x = 2 x_m 
= 2 \sqrt{a/b} \simeq 20 \, \mu m$.
These values are all compatible with the experiments already performed in 
Refs.~\cite{coh05, ACohen}.

\emph{Conclusion.}
We have considered an overdamped Brownian particle in a double--well potential as a generic model of a one--bit memory. We have investigated the probability distribution of the work and the heat dissipated during an erasure process and demonstrated that full erasure may be reached by dissipating an amount of heat below the Landauer limit.  We have shown that the occurrence of such events is exponentially suppressed, and therefore not observable, in macroscopic systems.  They play, however, an essential role in nanosystems and we have discussed a single--particle experiment where our predictions can be tested. Our main conclusion is that for small systems in general -- and Maxwell's demon in particular -- the macroscopic formulation of Landauer's principle does not hold, but has to be generalized to include heat fluctuations.

\begin{acknowledgments}
This work was supported by  the Emmy Noether Program of the DFG 
(Contract LU1382/1-1) and the cluster of excellence 
Nanosystems Initiative Munich (NIM).
\end{acknowledgments}

\end{document}